\documentclass[pra,aps,twocolumn]{revtex4}
\usepackage{epsf}

\def\binom#1#2{{#1\choose#2}}

\begin{document}

\title
{Quantum noise and mixedness of a pumped dissipative non-linear
oscillator}

\author{Ji\v r\'\i\ Bajer$^1$, Adam Miranowicz$^{2}$,
and Mateusz Andrzejewski$^2$}

\address{$^{1}$Department of Optics, Palack\'{y} University,
17.~listopadu 50, 772~00 Olomouc, Czech Republic}
\address{$^{2}$Nonlinear Optics Division, Physics Institute, Adam
 Mickiewicz University, 61-614 Pozna\'n, Poland}

\date{\today}

\begin{abstract}
Evolutions of quantum noise, characterized by quadrature squeezing
parameter and Fano factor, and of mixedness, quantified by quantum
von Neumann and linear entropies, of a pumped dissipative
non-linear oscillator are studied. The model can describe a signal
mode interacting with a thermal reservoir in a parametrically
pumped cavity with a Kerr non-linearity. It is discussed that the
initial pure states, including coherent states, Fock states, and
finite superpositions of coherent states evolve into the same
steady mixed state as verified by the quantum relative entropy and
the Bures metric. It is shown analytically and verified numerically
that the steady state can be well approximated by a nonclassical
Gaussian state exhibiting quadrature squeezing and sub-Poissonian
statistics for the cold thermal reservoir. A rapid increase is
found in the mixedness, especially for the initial Fock states and
superpositions of coherent states, during a very short time
interval, and then for longer evolution times a decrease in the
mixedness to the same, for all the initial states, and relatively
low value of the nonclassical Gaussian state.

\vspace{3mm} \noindent {\bf Keywords}: quantum entropy, quadrature
squeezing, sub-Poissonian statistics, Wigner function, Husimi
functions, Kerr non-linearity, Schr\"odinger cats, steady states.

\end{abstract}

\maketitle

\section{Introduction}

Quantum noise and mixedness are properties central for quantum
theory \cite{neumann}. It is obvious that a pure signal state
interacting with a thermal environment loses its purity and turns
into a mixed state. Since real optical devices suffer from losses,
it is important to study their influence on the noise and
mixedness during the state evolution.

In the case of interactions of a small number of the signal and
reservoir modes, the noise and mixedness parameters periodically
rise and fall. But in the case of the signal interacting with an
infinite number of the reservoir modes, the evolution is
irreversible and after a few characteristic time intervals, the
signal mode transforms into an asymptotic steady state depending on
external parameters of the system, e.g., an external classical
pump. One can show that when the pump is turned off, final steady
state of the signal will be the pure vacuum state, but when the
pump is turned on, the steady state is not the vacuum but a mixed
nonclassical state. We show further that the steady state can be
properly approximated by a nonclassical Gaussian state (NCGS).

In the next sections we will investigate a particular example of a
non-linear interaction, i.e., a nontrivial dynamics of initial
signal modes in a pumped resonator with a non-linear Kerr medium
when losses are included (see, e.g.,
\cite{drummond1,drummond2,kheru}). We are interested mainly in the
quantum noise and mixedness (purity) evolutions of the signal mode.
We discuss three different types of the signal initial states
including coherent states, Fock states and superpositions of
coherent states. We study the influence of the losses, pump and
non-linearity on the signal evolution, in particular the fast
decoherence of the initial coherent superpositions.

\section{Interaction model}

A pumped non-linear oscillator can be described in the interaction
picture by the following Hamiltonian \cite{drummond1,drummond2}:
\begin{eqnarray}
\hat{H}&=&\hat{H}_{I}+\hat{H}_{R}, \label{N01} \\
\hat{H}_I&=&\mathrm{i}( p\hat{a}^{\dag }-p^{\ast }\hat{a})
+G\hat{a} ^{\dag 2}\hat{a}^{2}. \label{N02}
\end{eqnarray}
where $\hat{a}$ and $\hat{a}^{\dag }$ are the annihilation and
creation operators of the signal mode, respectively. The first term
in Hamiltonian $\hat{H}_I$ describes a pump process with the
complex amplitude $p$, the second term represents a Kerr process
with the non-linear interaction constant $G$ and $\hat{H}_{R}$ is
the term describing the interaction of the signal mode with the
cold thermal reservoir. For simplicity we use the units where
$\hbar =1$. The model can be used in a description of, e.g., a
parametrically pumped cavity with a Kerr non-linearity. We adopt
the Heisenberg-Langevin approach \cite{perina}, in which the
Heisenberg equation for the signal mode reads is
\begin{equation}
\frac{\mathrm{d}\hat{a}}{\mathrm{d}t}=p-2\mathrm{i}G\hat{a}^{\dag
}\hat{a} ^{2}-\gamma _{0}\hat{a}+\hat{\cal L}, \label{N03}
\end{equation}
where $\hat{\cal L}$ is the Langevin force and $\gamma _{0}$ is the
loss parameter. For the optical modes at room temperature, the
number of thermal photons can be assumed negligible. Therefore, in
the following we assume the zero-temperature approximation. In that
case the standard rules hold for the Langevin force:
\begin{eqnarray}
\langle \hat{\cal L}( t_{1}) \hat{\cal L}( t_{2}) \rangle &=&
\langle \hat{\cal L}^{\dag }( t_{1}) \hat{\cal L}( t_{2}) \rangle
=\langle \hat{\cal L}( t_{1}) \rangle=0,
\nonumber \\
\langle \hat{\cal L}( t_{1}) \hat{\cal L}^{\dag }( t_{2}) \rangle
&=&2\gamma _{0}\delta ( t_{1}-t_{2}). \label{N04}
\end{eqnarray}
For the linear interaction ($G=0$) of the signal mode, an
analytical solution can be found. For the coherent initial state $|
\alpha \rangle $, the solution is $| \psi \rangle =| \alpha \left(
t\right) \rangle$ with the time-dependent complex amplitude
\begin{equation}
\alpha \left( t\right) =\alpha \mathrm{e}^{-\gamma
_{0}t}+\frac{p}{\gamma _{0}}\left( 1-\mathrm{e} ^{-\gamma
_{0}t}\right) . \label{N05}
\end{equation}
So the initial coherent state with the amplitude $\alpha $ finally
evolves into another coherent state with the amplitude $p/\gamma
_{0}$. The evolution is essentially finished during a short period
equal to the first few characteristic times $\tau =1/\gamma _{0}$.
For the model without a pump, the initial state turns into the
vacuum state.

In the non-linear Kerr interaction case, the evolution cannot be
solved analytically for arbitrary evolution times and suitable
numerical methods have to be used. The state evolution in this case
is much more complex and is the main subject of our study. Under
reasonable assumptions some analytical approximations can be
derived as well. After presenting our precise numerical results, we
will also give an exact quantum steady-state solution and its
semiclassical approximation in the long-time limit.

Alternatively we can start from the Liouville equation, which leads
to the following master equation
\begin{equation}
\frac{\mathrm{d}\hat{\rho}}{\mathrm{d}t}=\frac{1}{\mathrm{i}\hbar
}[\hat{H}, \hat{\rho}]=\frac{1}{\mathrm{i}\hbar
}[\hat{H}_{I},\hat{\rho}]+\hat{R}, \label{N06}
\end{equation}
where $\hat{R}$ for the zero temperature reservoir stands for
\begin{equation}
\hat{R}=\frac{1}{i\hbar }[\hat{H}_{R},\hat{\rho}]=\gamma
_{0}\left( 2\hat{a} \hat{\rho}\hat{a}^{\dag }-\hat{a}^{\dag
}\hat{a}\hat{\rho}-\hat{\rho}\hat{a} ^{\dag }\hat{a}\right) .
\label{N07}
\end{equation}
Master equation (\ref{N06}) can simply be expressed in the Fock
basis with the density matrix in the form
\begin{equation}
\hat{\rho}=\sum_{nm}\rho _{mn}|m\rangle \langle n| \label{N08}
\end{equation}
leading to a set of ordinary linear differential equations for the
density matrix elements $\rho _{mn}=\langle m|\hat{\rho}|n\rangle$.
For weak interaction fields the differential equations can be
solved numerically. The exact quantum results presented in this
article are obtained by applying this method.

To visualize evolution of quantum states (pure or mixed) generated
in our system, we apply the Husimi and Wigner functions, which are
the special cases of the Cahill-Glauber $s$-parametrized
quasidistribution function ${\cal W}^{(s)}(\beta)$ defined for $-1
\le s\le 1$ as follows \cite{cahill}:
\begin{eqnarray}
  {\cal W}^{(s)}(\beta) &=& \frac{1}{\pi} \,{\rm Tr}\,\{ \hat{\rho}\,
  \hat{T}^{(s)}(\beta)\},
\label{N09}
\end{eqnarray}
where $\hat{\rho}$ is the density matrix of the field and
\begin{eqnarray}
  \hat{T}^{(s)}(\beta) &=& \frac{1}{\pi} \int
  \exp(\beta\xi^*-\beta^*\xi) \hat{D}^{(s)}(\xi) \,{\rm d}^2 \xi,
\label{N10}
\end{eqnarray}
and $\hat{D}^{(s)}(\xi) = {\rm e}^{s|\xi|^2/2} \hat{D}(\xi)$ with
$\hat{D}(\xi)$ being the displacement operator. The
quasidistribution ${\cal W}^{(s)}(\beta)$ in the number-state basis
can be calculated as \cite{cahill}:
\begin{eqnarray}
{\cal W}^{(s)}(\beta) &=& \frac{1}{\pi}\sum_{m,n}\rho_{mn} \langle
n|\hat{T}^{(s)}(\beta) |m\rangle, \label{N11}
\end{eqnarray}
where
\begin{eqnarray}
\langle n|\hat{T}^{(s)}(\beta)|m\rangle = \left(\frac{n!}{m!}
\right)^{1/2} \left( \frac{2}{1-s}\right)^{m-n+1} \left(
\frac{s+1}{s-1} \right)^n \nonumber \\
\hspace{4mm}\times\; (\beta^*)^{m-n} \exp
\left(-\frac{2|\beta|^2}{1-s} \right) L_{n}^{m-n}\left(
\frac{4|\beta|^2}{1-s^2}\right) \nonumber \\ \label{N12}
\end{eqnarray}
given in terms of the associate Laguerre polynomials
$L_n^{m-n}(x)$. Eq. (\ref{N12}) for $s\rightarrow -1$ goes into the
simple expression
\begin{eqnarray}
\langle n|\hat{T}^{(-1)}(\beta )|m\rangle &=&\exp \left( -|\beta
|^{2}\right)\frac{\beta^n (\beta^*)^m}{ \sqrt{n!m!}} \label{N13}
\end{eqnarray}
as can be derived by observing that $\lim_{\epsilon \rightarrow 0}
\epsilon^n L_n^{m-n}(\frac{y}{\epsilon})=(-y)^n/n!$. Eq.
(\ref{N12}) can also be applied for $s\rightarrow 1$ if the limit
is taken carefully (see e.g. \cite{tanas}). The special cases of
${\cal W}^{(s)}(\beta)$ for $s=-1,0,1$ are known as the Husimi
$Q$-function, the Wigner $W$-function and Glauber-Sudarshan
$P$-function, respectively. For example, the $s$-parametrized
quasidistribution function for coherent state $| \alpha \rangle $
is given by the Gaussian distribution ${\cal W}^{(s)}( \beta) =
2/[\pi(1-s)] \exp[-2| \beta -\alpha | ^{2}/(1-s)]$, which for $s=1$
becomes Dirac's delta $\delta( \beta -\alpha)$. If a given state is
described by the $P$-function, which is positive definite and no
more irregular than Dirac's delta, then the state is classical,
otherwise the state is considered to be nonclassical. We use this
criterion to show that the steady state of our system is
nonclassical.

\section{Measures of quantum noise and mixedness}

Quantum noise properties of nonclassical light can be analysed in
terms of the Fano factor and quadrature noise variances. The first
parameter corresponds to direct photo-pulse detections, while the
second is related to homodyne detection schemes. The Fano factor is
defined by
\begin{equation}
F=\frac{\langle n^{2}\rangle -\langle n\rangle ^{2}}{ \langle
n\rangle }. \label{N14}
\end{equation}
which for Poissonian states satisfies $F=1$, while for sub- and
super-Poissonian states is $F<1$ and $F>1$, respectively. In
particular, the Fock states are sub-Poissonian with $F=0$
independently of the number of photons, coherent states are
Poissonian ($F=1$), while thermal chaotic states are
super-Poissonian since $ F=1+2\langle n\rangle>1$. The quadrature
noise squeezing parameter $S$ is defined as the minimum variance
$S=\min_{\theta }\langle (\Delta X_{\theta })^{2}\rangle$ over all
possible values of phase $\theta \in \left( 0,2\pi \right)$ of the
general quadrature operator $X_{\theta
}=\hat{a}\mathrm{e}^{-\mathrm{i}\theta }+\hat{a}^{\dag }\mathrm{e}
^{\mathrm{i}\theta }$.  For a coherent state $S=1$, while a state
with $S<1$ is referred to as the quadrature squeezed light, since
it has lower noise level than the coherent or vacuum state.

The most natural measure of mixedness of a state, given by
$\hat{\rho}$, is the von Neumann entropy \cite{neumann}
\begin{equation}
E=-\langle \ln \hat{\rho}\rangle =-\mathrm{Tr}\{\hat{\rho}\ln
\hat{\rho}\}. \label{N15}
\end{equation}
Density matrix $\hat{\rho}$ of any mixed state can be expressed as
the incoherent sum
\begin{equation}
\hat{\rho}=\sum_{k=1}^{N}p_{k}|  \psi _{k}\rangle \langle \psi
_{k}|  ,  \label{N16}
\end{equation}
of the orthogonal pure states $|\psi _{k}\rangle$, where $p_{k}$
are their weight factors being eigenvalues of the density matrix
$\hat{ \rho}$. As follows from the general properties of the
density matrix, it holds $ 0\leq p_{k}\leq 1$ and
$\sum_{k=1}^{N}p_{k}=1$. Thus, the von Neumann entropy can be
expressed as
\begin{equation}
E=-\sum_{k=1}^{N}p_{k}\ln p_{k}\geq 0. \label{N17}
\end{equation}
Another useful measure of mixedness is the linear entropy $L$
defined as
\begin{equation}
L=1-\mathrm{Tr}\{\hat{\rho}^2\}, \label{N18}
\end{equation}
where the second term in (\ref{N18}) is referred to as the purity
$P$ of the state
\begin{equation}
P=\mathrm{Tr} \{\hat{\rho}^2\}= \sum_{k=1}^{N}p_{k}^{2}.
\label{N19}
\end{equation}
The mixedness $L$ and the purity $P$ are complementary in the sense
that whenever the entropy increases the purity falls. For
completeness, we note other generalized measures of the mixedness
including (see, e.g., \cite{zyczkowski}): the Renyi entropies
defined by $H_{q}=(\ln \mathrm{Tr}\{\rho ^{q}\})/\left( 1-q\right)$
($q=2,3,...$), the Neumann-Renyi entropy given by $H_{2}=-\ln
\mathrm{Tr}\{\hat{\rho}^{2}\}=-\ln P$ or the participation ratio
defined to be $R=1/\mathrm{Tr}\{\rho ^{2}\}=1/P$.

It is easy to show the following properties of the mixedness
parameters $E$ and $L$: For any pure state it holds $E=L=0$. For a
mixture of two orthogonal states it holds $0\leq L\leq 1/2$ and
$0\leq E\leq \ln 2$. For a balanced mixture of two orthogonal
modes $p_{1}=p_{2}=1/2$ it holds $L=1/2$ and $E=\ln 2$ as their
maxima. For a mixed state composed of $N$ orthogonal pure states
it holds $0\leq L\leq (N-1)/N$ and $0\leq E\leq \ln N$. For a
homogeneous superposition it is $p_{k}=1/N$ and therefore
$L=(N-1)/N$ and $E=\ln N$. So, for a state strongly mixed with the
reservoir, $L$ can increase to one and $E$ to infinity.

For thermal or chaotic states, the density matrix is diagonal with
\begin{equation}
\rho _{kk}=p_{k}=\frac{\langle n\rangle ^{k}}{\left( 1+\langle
n\rangle \right) ^{1+k}}, \label{N20}
\end{equation}
implying that the linear entropy is
\begin{equation}
L_{\mathrm{chaot}}=\frac{2\langle n\rangle}{1+2\langle n\rangle }
\label{N21}
\end{equation}
and the corresponding von Neumann entropy is
\begin{equation}
E_{\mathrm{chaot}}=-\ln \frac{\langle n\rangle ^{\langle n\rangle
}}{\left( 1+\langle n\rangle \right) ^{1+\langle n\rangle }}.
\label{N22}
\end{equation}
Here, $\langle n\rangle $ stands for the number of photons in the
chaotic mode. For the vacuum state $\langle n\rangle =0$, which is
a pure state, we get $E=L=0$, while for the chaotic state with $
\langle n\rangle =1$ we get $E=\allowbreak \ln 4\approx
\allowbreak 1.\,\allowbreak 386$ and $L=2/3$.

The thermal state is defined as the state, which maximizes the
entropy when at the same time the energy is fixed. So, the state
with a fixed number of photons $\langle n\rangle $ has the upper
limit of entropy and the following inequality holds $E\leq
E_{\mathrm{\max }}=E_{\mathrm{chaot}}$. The linear entropy has the
upper limit of
\begin{equation}
L_{\max }=1-\frac{1+2\langle n\rangle }{\left( 1+3\langle n\rangle
\right) \left( 1+3\langle n\rangle /2\right) } \label{N23}
\end{equation}
reached for a state with the descending arithmetic sequence type
distribution
\begin{equation}
p_{k}=\frac{2}{2+3\langle n\rangle }\left( 1-\frac{k}{ 1+3\langle
n\rangle }\right) . \label{N24}
\end{equation}
For example, the state with $\langle n\rangle =1$ has the upper
limit of the linear entropy equal to $L_{\max}=7/10$.

\section{Numerical analysis of mixedness and noise}

\subsection{Ideal non-linear oscillator without pump}

It is well known that in the case of Kerr dynamics without losses
and without a pump, the initial coherent state evolves
periodically into a nonclassical light with highly reduced quantum
noise \cite{squeezing} (see also \cite{bajer1} and references
therein) and also becomes superpositions of two (Schr\"odinger
cats) \cite{cats} and more (Schr\"odinger kittens) \cite{kittens}
coherent states. If the initial state is
\begin{equation}
|  \psi _{0}\rangle =\sum_{k=0}^{\infty }c_{k}| k\rangle ,
\label{N25}
\end{equation}
then its evolution is described by
\begin{equation}
|  \psi \rangle =\sum_{k=0}^{\infty }c_{k}\mathrm{e}^{-\mathrm{
i}k\left( k-1\right) Gt}| k\rangle . \label{N26}
\end{equation}
being clearly periodic with the time period $T=\pi /G$, as the
state is $| \psi \left( T\right) \rangle =|  \psi _{0}\rangle $.
For example, the coherent initial state $| \psi _{0}\rangle =|
\alpha \rangle $ at the time $t=T/2$ becomes the coherent
superposition
\begin{equation}
|  \psi _{2}\rangle = \frac{1}{\sqrt{2}}\left(
\mathrm{e}^{\mathrm{i}\pi /4}| \mathrm{i}\alpha \rangle
+\mathrm{e}^{-\mathrm{i}\pi /4}| -\mathrm{i}\alpha \rangle \right)
. \label{N27}
\end{equation}
More generally, at the evolution times $t=\frac{m}{n}T$, which is
a rational fraction of the period $T$, the output state is given
as a coherent superposition of $n$ coherent states dislocated
regularly on the circumference of a circle with a radius $| \alpha
| $. A typical example of such a Kerr evolution can be seen in
figure 1.
\begin{figure}
\epsfxsize=8.5cm\epsfbox{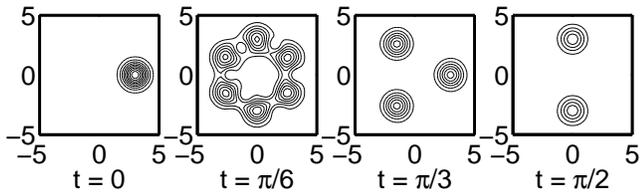} \caption{Time snapshots of
Husimi $Q$-function for the initial coherent state $|
\protect\alpha \rangle =|  3\rangle $. Kerr parameter $G=1$, no
loss $ \protect\gamma _{0}=0$ and no pump $p=0$.}
\end{figure}
\begin{figure}
\epsfxsize=8.5cm\epsfbox{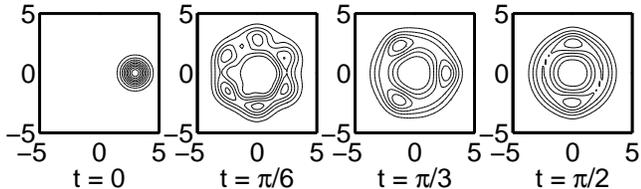} \caption{Same as in figure 1
but for the dissipative model with $\protect\gamma _{0}=1$.}
\end{figure}
Obviously, the initial pure state remains pure ($E=L=0$).

\subsection{Dissipative non-linear oscillator without pump}

When losses are involved in the system without a pump, the
evolution is no more periodic and it ends in the vacuum state.
During the time evolution, an initial state goes through various
nonclassical states and through strongly mixed states. Some
analytical solutions of the dissipative Kerr non-linear oscillator
have been obtained both for 'quiet' ($T\approx 0$) \cite{solution0}
and `noisy' ($T>0$) \cite{solution1} reservoirs.

The role of dissipation is clearly seen by comparing the evolution
of the initial coherent state $|  \alpha = 3\rangle $ without
(figure 1) and with (figure 2) losses. The parameters used in
figures 1 and 2 are $G=1$, $p=0$ and the loss parameter $\gamma
_{0}$ is 0 in figure 1 and $0.1$ in figure 2. Due to losses, no
superpositions of coherent states arise in the evolution presented
in figure 2. The effect can be explained as a fast destruction of
internal coherence.
\begin{figure}
\centerline{\epsfxsize=8cm\epsfbox{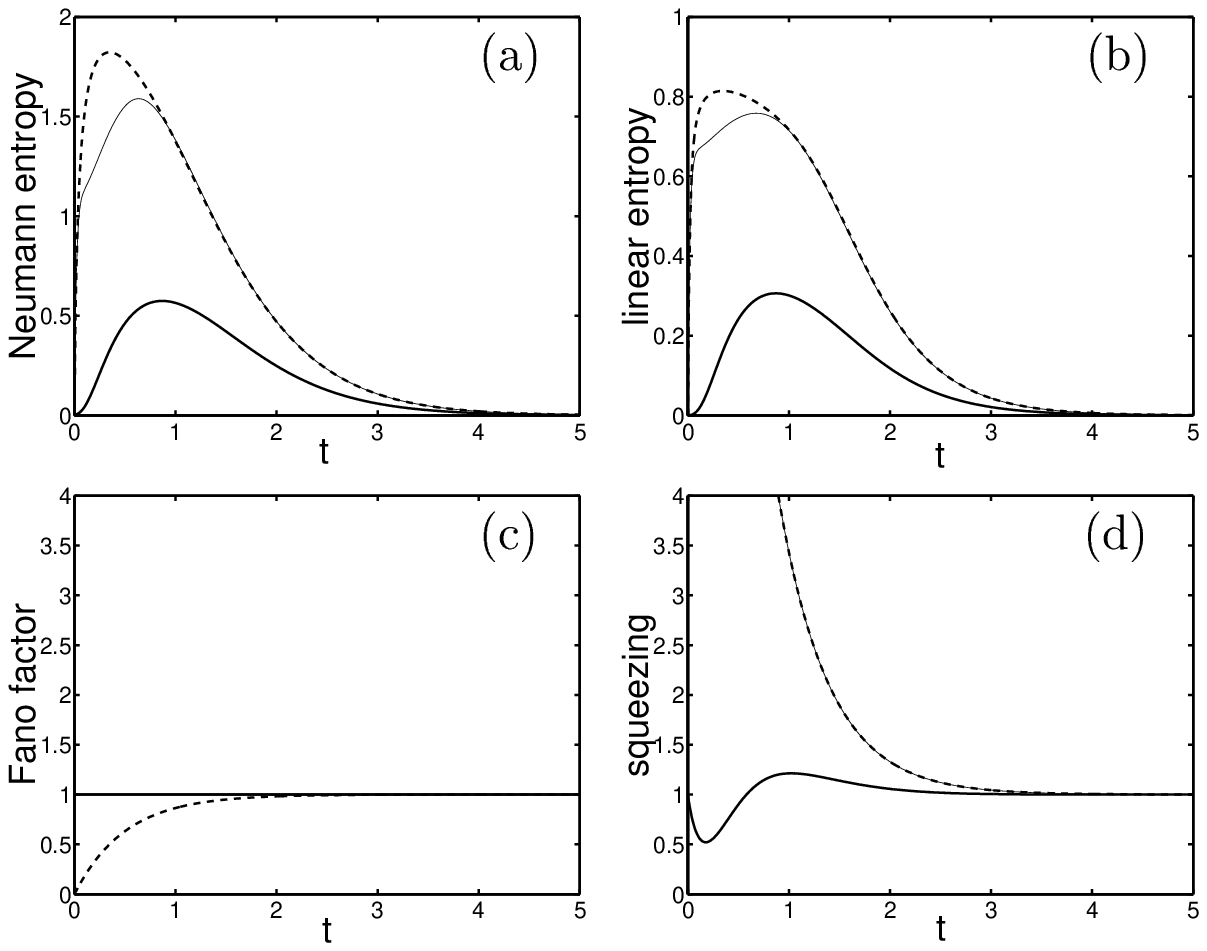}} \caption{Evolution
of the mixedness and noise for the system without a pump: (a) the
von Neumann entropy $E$, (b) linear entropy $L$, (c) Fano factor
$F$ and (d) squeezing parameter $S$ for the initial coherent state
$|\alpha=3\rangle $ (thick solid curves), the Fock state $
|n=9\rangle $ (dashed curves) and the superposition of three
coherent states $| \protect\psi _{3}\rangle$ (thin solid curves).
Resonator parameters are $p=0$, $G=0.2$ and $\protect\gamma
_{0}=1$.}
\end{figure}
\begin{figure}
\centerline{\epsfxsize=5cm\epsfbox{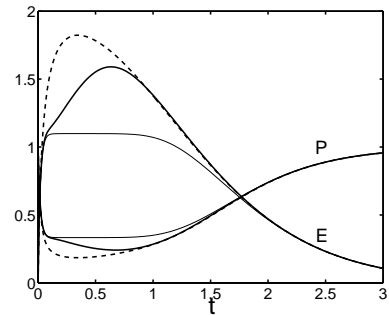}} \caption{Evolution
of the purity $P$ and entropy $E$ for the initial Fock state $|
n=9\rangle $ (dotted curve) and for the superposition of three
coherent states $| \protect\psi _{3}\rangle $ with Kerr (thick
solid) and without Kerr (thin solid curves). The resonator
parameters are same as in figure 3.}
\end{figure}
Now we will analyse the evolution of the mixedness, characterized
by the von Neumann ($E$) and linear ($L$) entropies, and the noise
in terms of the Fano factor $F$ and quadrature squeezing parameter
$S$ given in figure 3 for the parameters $G=0.2$, $p=0$ and $\gamma
_{0}=1$. First we focus on the evolution of the initial coherent
state $| \alpha \rangle =| 3\rangle $ depicted by thick solid
curves in figure 3. As seen, the signal state is maximally mixed at
the time $t\approx 0.87$ when the entropies reach the values of
$E\approx 0.57$ [figure 3(a)] and $L\approx 0.31$ [figure 3(b)].
During the evolution, the photocount statistics of the initial
coherent state remains Poissonian. It is a direct consequence of
the fact that the diagonal terms of the density matrix keep the
Poissonian character for all the times. It can be shown that it
holds exactly that
\begin{equation}
p\left( n\right) =\exp \left( -\langle n\rangle \right) \frac{
\langle n\rangle ^{n}}{n!}, \label{N28}
\end{equation}
where $\langle n\rangle =|  \alpha |  ^{2}\exp \left( -2\gamma
_{0}t\right)$, which explains why the Fano factor remains constant
and equal to one for all the times [see figure 3(c)]. On the other
hand, the quadrature noise evolves leading to the maximum
squeezing of $S\approx 0.52$ at the time $t\approx 0.18$. In
figure 3, we have also presented by thin solid curves the
evolutions of the mixedness and the noise of the initial
superposition of three coherent states (Schr\"{o}dinger kittens)
\begin{equation}
|  \psi _{3}\rangle = N_3 \left( | 3\rangle +|
3\mathrm{e}^{\mathrm{i}\frac{2\pi }{3} }\rangle +|
3\mathrm{e}^{-\mathrm{i}\frac{2\pi }{3} }\rangle \right),
\label{N29}
\end{equation}
where $N_3$ is a normalization constant. One can observe a rapid
increase in the mixedness measured by both $E$ in figure 3(a) and
$L$ in figure 3(b) during a very short time interval $t<0.05$. It
can be explained as a general effect of the coherence loss of
quantum components of the superposition $| \psi _{3}\rangle $. In
general, it can be shown that the superposition of three strong
coherent states (normalized by $N\approx 1/\sqrt{3}$):
\begin{equation}
|  \psi \rangle = N \left( |  \alpha _{1}\rangle +| \alpha
_{2}\rangle +| \alpha _{3}\rangle \right) \label{N30}
\end{equation}
exponentially fast evolves into the mixed state
\begin{equation}
\hat{\rho}= N^2 \left( |  \alpha _{1}\rangle \langle \alpha _{1}|
+| \alpha _{2}\rangle \langle \alpha _{2}|  +| \alpha _{3}\rangle
\langle \alpha _{3}|  \right) \label{N31}
\end{equation}
as a consequence of the reservoir influence. The characteristic
time of the decoherence process between $|  \alpha _{i}\rangle $
and $|  \alpha _{k}\rangle $ ($i \neq k$ for $i,k=1,2,3$) can be
estimated as
\begin{equation}
\tau _{ik}\approx \left( 2\gamma _{0}|  \alpha _{i}-\alpha _{k}|
^{2}\right) ^{-1}. \label{N32}
\end{equation}
So for strong fields, the decoherence process is much faster than
the dissipative process itself. The mixedness measured by the
linear entropy increases to a value of $L=2/3$, while by the von
Neumann entropy increases to $E=\ln 3\approx \allowbreak 1.10$. In
our case of $| \psi _{3}\rangle $ the increase in the linear
entropy can be approximated by
\begin{equation}
L\approx \frac 23 \left[ 1-\exp \left( -6\gamma _{0}|  \alpha |
^{2}t\right) \right] . \label{N33}
\end{equation}
whose validity is verified by the results shown in figures 3(a,b)
and 4. We note that a better agreement with the decoherence theory
is achieved for the system when the Kerr non-linearity is switched
off (corresponding to thin solid curves in figure 4) rather than is
switched on (depicted by thick solid curves). For longer times, the
noise dominates the process and the signal state turns into the
vacuum state. Thus, the squeezing parameter evolves into the value
of one as shown in figure 3(d). As for the initial coherent state,
the Fano factor for the initial superposition state $| \psi
_{3}\rangle $ remains unchanged and equal to unity [see figure
3(c)] during the evolution of the unpumped system for arbitrary
values of the interaction $G$ and loss $\gamma_0$ parameters.

For the initial Fock states $| n= 9\rangle $, a rapid increase in
the mixedness can also be observed in figure 3(a,b) as depicted by
dotted curves. For $t\approx 0.35$ we find the mixedness
parameters to be $L\approx 0.8$ and $E\approx 1.8$. The mechanism
of decoherence and increase in the mixedness differ from those for
the initial superposition of coherent states. We can explain the
observed values simply as follows: by solving directly the
Schr\"{o}dinger equation for the initial Fock state $| n\rangle $,
we get a diagonal density matrix at time $t$ described by the
binomial distribution
\begin{equation}
p\left( k\right) =\rho _{kk}=\binom{n}{k}\left(
\mathrm{e}^{-2\gamma _{0}t}\right) ^{k}\left(
1-\mathrm{e}^{-2\gamma _{0}t}\right) ^{n-k}, \label{N34}
\end{equation}
which is surprisingly completely independent of the Kerr
non-linearity $G$. The mixedness parameter $L=1-\sum p^{2}\left(
k\right) $ sums up to an expression proportional to a
hypergeometric function $_2F_1(-n,-n,1,x)$, which is difficult to
handle. But we can easily calculate the variance
\begin{equation}
\overline{\Delta ^{2}k}=n\left( \mathrm{e}^{-2\gamma _{0}t}\right)
\left( 1- \mathrm{e}^{-2\gamma _{0}t}\right) \label{N35}
\end{equation}
which reaches a maximum at the time moment when $\mathrm{e}^
{-2\gamma _{0}t}=1/2$. So we can estimate the searched maximum of
the linear entropy $L$ at this moment. We have
\begin{equation}
L_{\max}\approx 1-\sum_{k=0}^{n}\binom{n}{k}^{2}2^{-2n}=1- \frac{(
2n) !}{( 2^{n}n!) ^{2}} \label{N36}
\end{equation}
and for large $n$ we get the following simple approximation
\begin{equation}
L_{\max}\approx 1- \frac{1}{\sqrt{\pi n}}, \label{N37}
\end{equation}
by applying the Stirling formula $n!\approx \sqrt{2\pi
n}e^{-n}n^{n}$. Thus, we can analytically estimate the loss of
purity for $n=9$ as $ L_{\max }\approx 0.815$ according to
(\ref{N36}) (or slightly less accurate as 0.812 according to
(\ref{N37})), which can be expected at the time $t\approx 0.347$.
We can calculate the entropy of the state at the same time moment
to get $E\approx \allowbreak 1.\,\allowbreak 823$. These
estimations are in full agreement with the precise numerical
results given in figure 5(a,b). Finally, we note that the squeezing
parameter decreases from $S=19$, while  the Fano factor increases
from $F=0$ for the initial Fock state to reach the value of one for
the vacuum state in the time limit.

\subsection{Dissipative non-linear oscillator with pump}

Here, we will analyse the system with the classical pump switched
on, for which the initial state finally evolves into a stationary
non-vacuum state. Its properties are determined by the pump
intensity $p$ and by the passive parameters $ \gamma _{0}$ and $G$.
We will study three typical evolutions for different initial fields
including coherent states,  Fock states and superpositions of three
coherent states. Figure 5 shows the Wigner function evolution of
the coherent initial state $| \alpha =3\rangle $, figure 6 shows
the evolution of the Fock initial state $| n= 9\rangle $ and figure
7 shows evolution of the superposition of three coherent states,
given by (\ref{N29}), for the same resonator parameters $p=5$,
$\gamma _{0}=1$ and $G=0.2$. All the three states have been
selected to have the same initial energy $\langle n\rangle =9$ (in
the last case approximately). Six time snapshots of the Wigner
function of the signal state are presented for the time moments
$t=0,1/3,2/3, 1,5$ and for a very long time, practically
corresponding to $\infty$. In figure 5, the coherent initial state
rotates and evolves in a banana-shape state and finally ends in a
steady state. In figure 6, the initial Fock state of a ring shape
is deformed and squeezes before ending in a steady state. In figure
7, the initial superposition of the coherent states rotates,
deforms and its components converge into the same steady state.
\begin{figure}
\centerline{\epsfxsize=8cm\epsfbox{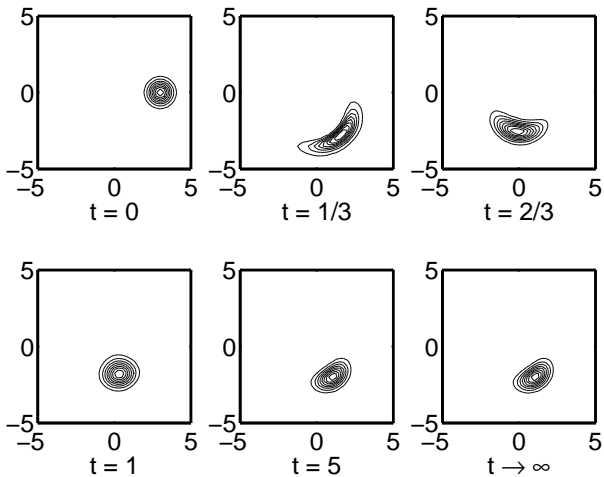}} \caption{Time
snapshots of Wigner function for the initial coherent state $|
\alpha=3\rangle $. Resonator parameters are $G=0.2$, $\gamma
_{0}=1$ and $p=5$.}
\end{figure}
\begin{figure}
\centerline{\epsfxsize=8cm\epsfbox{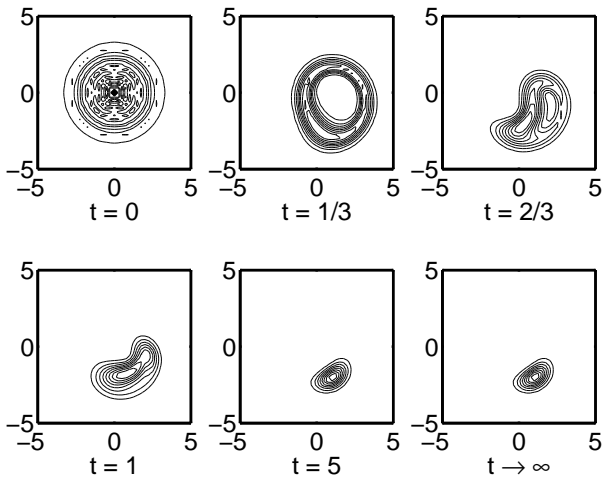}} \caption{Same as in
figure 5 but for the initial Fock state $|n=9\rangle $.}
\end{figure}
\begin{figure}
\centerline{\epsfxsize=8cm\epsfbox{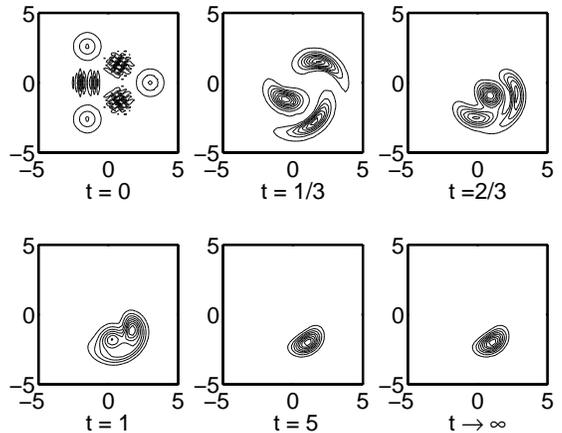}} \caption{Same as in
figure 5 but for the initial coherent state superposition $|
\protect\psi _{3}\rangle $.}
\end{figure}
\begin{figure}
\centerline{\epsfxsize=8cm\epsfbox{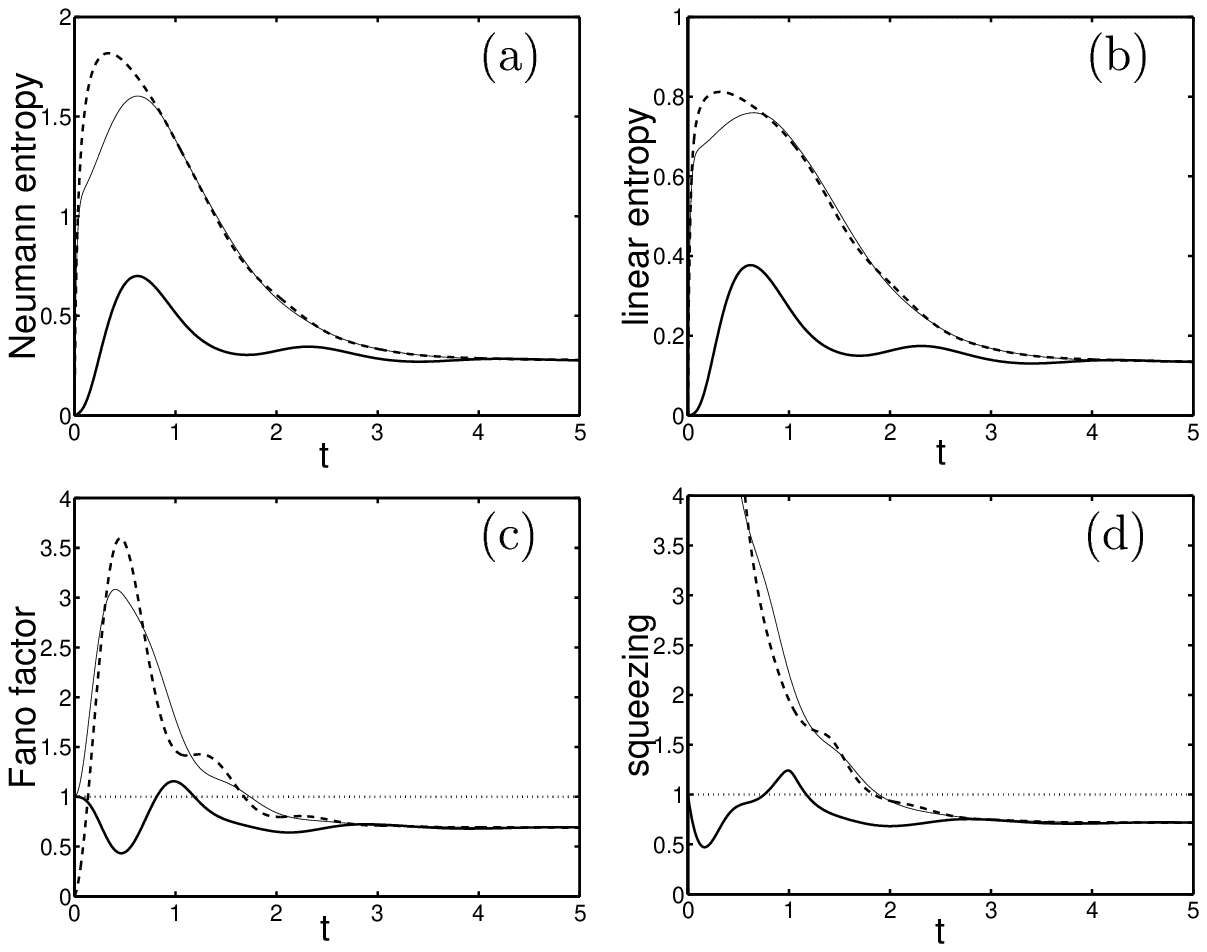}} \caption{Evolution
of the mixedness and noise parameters same as in figure 4 but for
the pumped system with $p=5$.}
\end{figure}
For the coherent initial state, the phase-space evolutions of the
mean quantum $\langle \hat{a}(t)\rangle $ and  classical
$\alpha(t) $ amplitudes in the time interval $0<t<10$ are
presented in figure 9. Both the spirals start at the point $\alpha
=3$, and end at $\alpha \approx 1-2\mathrm{i}$. Note that the
evolutions are very similar to each other even for the other
times, which shows the validity of our semiclassical
approximation.
\begin{figure}
\centerline{\epsfxsize=5cm\epsfbox{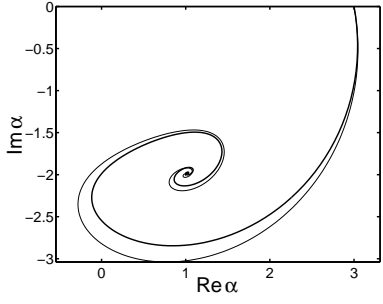}} \caption{Phase space
trajectory of quantum $\langle a\left( t\right) \rangle $ (thick
curve) and classical $\protect\alpha \left( t\right) $ (thin curve)
solutions for $t\in \left( 0,10\right) $.}
\end{figure}
The evolutions of the mixedness parameters $E$ and $L$ and the
quantum noise parameters $F$ and $S$ for all the three initial
states in the pumped system are depicted in figure 8. Note a rapid
increase in the mixedness especially in the cases of the initial
Fock state and a superposition of coherent states for very short
evolution times, and then for longer times, a decrease in the
mixedness for all the initial states. We observe that the
evolutions for short times are similar to those presented in
figure 3 for the unpumped system. However, for longer times (even
for $t>5$ for the parameters of figure 8) all the three initial
states of the pumped system evolve into some mixed steady state
different from the vacuum state. This is in contrast to the
evolution of the unpumped system for which $E(p=0)$  and $L(p=0)$
approach zero, while $S(p=0)$ and $F(p=0)$ approach one in the
time limit (see figure 3), since the dissipative system without
external pumping evolves into the vacuum state for any initial
fields.
\begin{figure}
\centerline{\epsfxsize=7cm\epsfbox{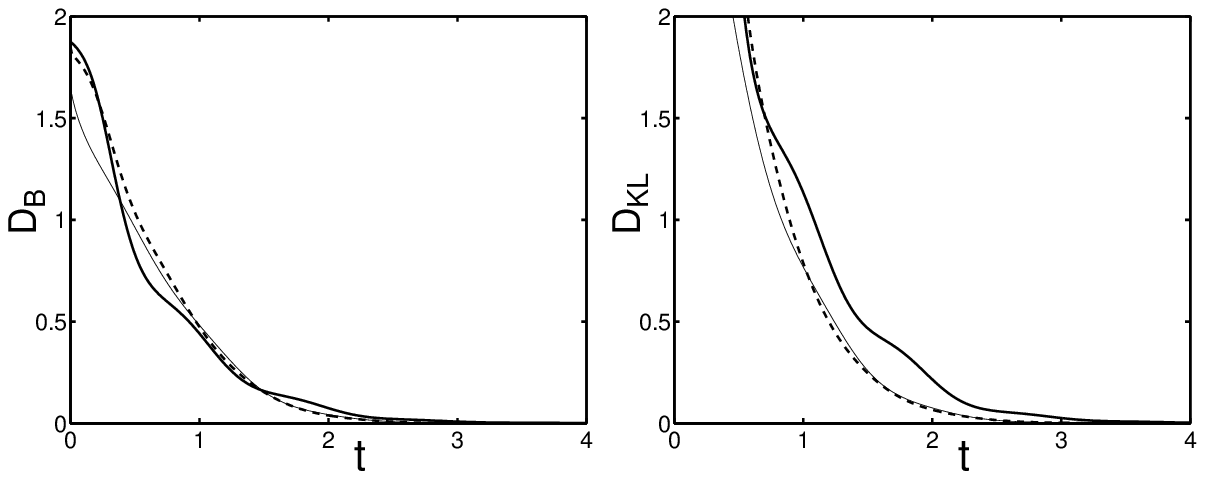}} \caption{Evolution
of the Bures metric ($D_B$) and the quantum relative entropy
($D_{K\!L}$)  for the same initial states as in figure 8.}
\end{figure}
It is worth noting, although it is out of the focus of our
interest, that the weakly-pumped non-linear oscillator ($p\ll G$)
enables the so-called optical state truncation \cite{leonski} and
can be used for optical qubit generation.

\subsection{Steady-state solution}

In figures 5--7 we have presented the evolutions of the Wigner
functions up to the time $t=5$ and then checked for much longer
times, practically, for $t\rightarrow\infty$. From this study we
can conclude that all the three initial states end in the same
steady state, which is centred around the same value of the complex
amplitude $\alpha \approx 1-2\mathrm{i}$. All the other parameters
studied that is: $E\approx 0.278$, $L\approx 0.135$, $\langle
n\rangle \approx 5.13$, $F\approx 0.69$ and $S\approx 0.72$ of the
steady state are the same as well. So, the steady-state solution is
apparently unique and totally independent of the initial states. It
can be therefore fully determined by the resonator parameters only.
While for the system without a pump the evolution ends in the
vacuum state, for the pumped system the asymptotic state is neither
the vacuum nor pure, and can have intriguing noise properties. We
can prove it by a decomposition of the steady-state density matrix
in its spectrum of eigenstates.

Drummond and Walls found that no Glauber-Sudarshan $P$-function
exists in the steady-state except as a generalized function
\cite{drummond1}. The latter was used by Kheruntsyan to find the
following explicit form of the steady-state density matrix
\cite{kheru}
\begin{equation}
\hat{\rho}_{\rm ss} =C\sum_{n,m}\frac{(\epsilon ^{*})^m\epsilon
^{n}}{\sqrt{m!n!}}\frac{ _{0}F_{2}(\lambda ^{\ast }+m,\lambda
+n,|\epsilon |^{2})}{\Gamma (\lambda ^{\ast }+m)\Gamma (\lambda
+n)}|n\rangle \langle m| \label{N39}
\end{equation}
where $C=\Gamma (\lambda ^{\ast })\Gamma (\lambda )/_{0}F_{2}
(\lambda ^{\ast },\lambda ,2|\epsilon |^{2})$, $\epsilon =-ip/G$,
$\lambda =-i\gamma _{0}/G$, and $_{0}F_{2}$ is the generalized
Gauss hypergeometric function. Eq. (\ref{N39}) readily enables
calculation of the moments $\langle (\hat{a}^{\dag
})^{m}\hat{a}^{n}\rangle_{\rm ss}$ by the expressions summing up to
\begin{equation}
\langle (\hat{a}^{\dag })^{m}\hat{a}^{n}\rangle_{\rm ss} =C
(\epsilon ^{*})^m \epsilon ^{n}\,\frac{_{0}F_{2}(\lambda ^{\ast
}+m,\lambda +n,2|\epsilon |^{2})}{\Gamma (\lambda ^{\ast
}+m)\Gamma (\lambda +n)} \label{N40}
\end{equation}
which corresponds to a slightly modified Drummond-Gardiner formula
\cite{drummond2}. Solution (\ref{N39}), with the help of
definition (\ref{N11}), also enables calculation of the
steady-state Husimi and Wigner functions. E.g., the latter can be
given by \cite{kheru}
\begin{equation}
{\cal W}^{(0)}_{\rm ss}(\beta )=N^{(0)}e^{-2|\beta |^{2}}\left|
\frac{J_{\lambda -1}(\sqrt{ -8\epsilon \beta ^{\ast }})}{(\beta
^{\ast })^{(\lambda -1)/2}}\right| ^{2} \label{N41}
\end{equation}
where $J_{\lambda -1}(x)$ is the Bessel function and $N^{(0)}$ is
the normalization constant.

To find a physical insight of the steady-state density matrix
(\ref{N39}),  we rewrite it as the incoherent superposition
(\ref{N16}):
\begin{equation}
\hat{\rho}_{\rm ss}=p_{0}|  \psi _{0}\rangle \langle \psi _{0}|
+p_{1}| \psi _{1}\rangle \langle \psi _{1}| +p_{2}| \psi
_{2}\rangle \langle \psi _{2}| +... \label{N42}
\end{equation}
of the orthogonal pure states $|  \psi _{k}\rangle $ with the
weight factors $p_{k}$, which can be found numerically. Applying
the eigenvalue method for the steady state we have obtained all
the spectrum of the final state. Wigner functions of the first
three most important components are displayed in figure 11. The
weight coefficients of the strongest components are in the
sequence of $p_{0}\approx 0.928$, $p_{1}\approx 0.068$ and
$p_{2}\approx 0.004$. The first component $|  \psi _{0}\rangle $,
apart from the second and third components, is strongly
nonclassical with its quadrature noise of about $S_{0}\approx
0.60$, while the compound steady state has the quadrature noise
level just about $S\approx 0.72$.
\begin{figure}
\centerline{\epsfxsize=9cm\epsfbox{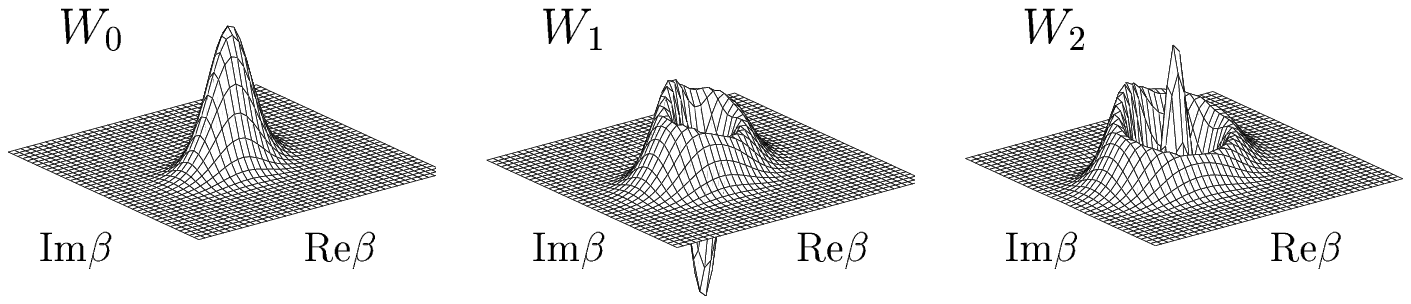}} \caption{Wigner
functions $ W_j={\cal W}_{\rm ss}^{(0)}(| \protect\psi
_{j}\rangle)$  ($j=0,1,2$) of the first three pure state
components of the steady state for the resonator parameters
$G=0.2,\;p=5,\;\protect\gamma _{0}=1$.}
\end{figure}
To verify the conclusion that the density matrices $\hat{\rho}(t)$
for different initial conditions evolve into the same steady state
described by the density matrix (\ref{N39}), we calculate the Bures
metric and the quantum relative entropy (the generalized
Kullback-Leibler distance) between $\hat{\rho}(t)$ and
$\hat{\rho}_{\rm ss}$ defined as follows \cite{vedral}
\begin{eqnarray}
D_B\{\hat{\rho}(t)||\hat{\rho}_{\rm ss}\} &=& 2-2\,{\rm
Tr}\{[\sqrt{\hat{\rho}_{\rm ss}}\hat{\rho}(t)
\sqrt{\hat{\rho}_{\rm ss}}]^{1/2}\}, \label{N43} \\
D_{K\!L}\{\hat{\rho}(t)||\hat{\rho}_{\rm ss}\} &=& {\rm Tr}
\{\hat{\rho}(t)[\ln\hat{\rho}(t)-\ln\hat{\rho}_{\rm ss}]\},
 \label{N44}
\end{eqnarray}
respectively. The square of trace in (\ref{N43}) is the so-called
Uhlmann transition probability or the Jozsa fidelity for mixed
states \cite{jozsa}. Figure 10 clearly shows that the Bures metric
and quantum relative entropy approach zero for the three different
initial fields including coherent states, their superpositions and
the Fock states.

In the next section we will show that the steady state can be well
described as a nonclassical Gaussian state at least for high-pump
intensities.

\section{Linearized approximation of the steady-state solution}

Now we will give a linearized approximation of the steady-state
solution of the Heisenberg equation (\ref{N03}).  We start with
the classical approximation for strong fields, when the quantum
noise can be neglected and the signal mode can be described by a
complex amplitude. Heisenberg equation (\ref{N03}) gives the
following ordinary differential equation
\begin{equation}
\frac{\mathrm{d}\alpha }{\mathrm{d}t}=p-2\mathrm{i}G|  \alpha |
^{2}\alpha -\gamma _{0}\alpha  \label{N45}
\end{equation}
for the complex amplitude $\alpha $. Minor differences between the
exact quantum mean amplitude $\langle a\left( t\right) \rangle $
and the classical amplitude $\alpha \left( t\right) $ for the
coherent initial state $|  \alpha = 3\rangle $ and for the
resonator parameters $p=5$, $G=0.2$ and $\gamma _{0}=1$ can be
seen in the phase-space trajectory depicted in figure 9.

If we are interested in the steady-state solution only, we may
replace the left hand side of equation (\ref{N45}) by zero and
solve it. The complex equation can be transformed into the real
equation
\begin{equation}
|  p|  ^{2}=\left( \gamma _{0}^{2}+4G^{2}| \alpha | ^{4}\right) |
\alpha |  ^{2}, \label{N46}
\end{equation}
which is a cubic equation in intensity $|  \alpha | ^{2}$. A
simple analysis shows that the amplitude $| \alpha | $ of the
steady-state solution is a monotonic function of the pump
parameter $| p| $ and so equation (\ref{N46}) has a unique
solution and no threshold can be observed. Since the explicit form
of the solution is too complex to find some relevant physics in
it, we will not present it here. We note only that for the
resonator parameters $p=5$, $G=0.2$ and $\gamma _{0}=1$ assumed in
the examples investigated in the previous section one can find the
exact classical steady-state solution $\allowbreak |  \alpha |
^{2}=5$ and $\alpha =1-2\mathrm{i}$. In particular, (i) for $p=0$,
equation (\ref{N46}) has a trivial solution of $\alpha =0$, (ii)
for weak pump $| p| ^{2}\ll \gamma _{0}^{3}/2G$, the amplitude
grows linearly with the pump intensity as $\alpha \approx p/\gamma
_{0}$ and (iii) for very strong pump $| p| ^{2}\gg \gamma
_{0}^{3}/2G$, the amplitude grows with the third root of the pump
intensity
\begin{equation}
\alpha \approx -\mathrm{i}\frac{p}{|  p|  }\left( \frac{ | p|
}{2G}\right) ^{1/3}. \label{N47}
\end{equation}
To describe the noise properties of the signal mode we will search
a solution to equation (\ref{N03}) as a sum of two components
$\hat{a}=\alpha +\hat{A}$, where $\alpha$ is the strong classical
complex amplitude and $\hat{A}$ is the weak quantum noise
operator. By inserting it into equation (\ref{N03}) and after
linearization one gets the following equation for the noise
operator
\begin{equation}
\frac{\mathrm{d}\hat{A}}{\mathrm{d}t}=-\gamma \hat{A}-\delta
\hat{A}^{\dag }+ \hat{\cal L},  \label{N48}
\end{equation}
where the complex coefficients $\gamma $ and $\delta $ depend on
time through the amplitude $\alpha $ and are given by the
relations
\begin{equation}
\gamma =\gamma _{0}+4\mathrm{i}G|  \alpha |  ^{2}, \qquad \delta
=2\mathrm{i}G\alpha ^{2}. \label{N49}
\end{equation}
The solution of equation (\ref{N48}) is a Gaussian state, defined
by (\ref{N55}) in Appendix A, exhibiting purely nonclassical
properties, thus referred to as the NCGS. The time evolution of
the NCGS parameters $B$ and $C$ is described by the two
differential equations
\begin{eqnarray}
\dot{B} &=&-2\left( \gamma +\gamma ^{\ast }\right) B-\left( \delta ^{\ast
}C+\delta C^{\ast }\right) , \nonumber \\
\dot{C} &=&-\delta \left( 1+2B\right) -2\gamma C. \label{N50}
\end{eqnarray}
From them one can obtain numerically the linearized solution of the
Heisenberg equation (\ref{N03}), but it is meaningless when we know
the exact quantum solution. Instead of that we will focus on the
steady-state solution only. Since $|\gamma|>|\delta|$, a stationary
solution of the operator equation (\ref{N48}) exists. It is the
NCGS with the following parameters \cite{bajer2}
\begin{equation}
2B=\frac{|  \delta |  ^{2}}{|  \gamma | ^{2}-| \delta |
^{2}},\qquad 2C=- \frac{\delta \gamma ^{\ast }}{| \gamma | ^{2}-|
\delta | ^{2}}. \label{N51}
\end{equation}
Since we know all the three parameters $\alpha $, $B$ and $C$ of
the NCGS, we can simply estimate the other parameters of the steady
state, where the corresponding formulas are given in Appendix A.
For example, for the quadrature noise squeezing parameter
(\ref{N60}) one can derive a simple formula
\begin{equation}
S=\frac{|  \gamma |  }{|  \gamma | +| \delta | }\leq 1 \label{N52}
\end{equation}
demonstrating that the steady state is squeezed. The parameter $x$,
defined by (\ref{N57}), is explicitly given by
\begin{equation}
x=\frac{1}{2}\left( \sqrt{\frac{|  \gamma |  ^{2}}{ | \gamma |
^{2}-|  \delta |  ^{2}}} -1\right) . \label{N53}
\end{equation}
Numerically, for the resonator parameters $p=5$, $G=0.2$ and
$\gamma _{0}=1$, we get $\alpha =1-2\mathrm{i}$, $\gamma
=1+4\mathrm{i}$, $\delta =1.\,\allowbreak 6-1.\,\allowbreak
2\mathrm{i}$, $S\approx 0.673$, $F\approx 0.711$, $x\approx 0.072$,
$p_{0}\approx 0.933\,$, $p_{1}\approx 0.06\allowbreak 2$ and
$p_{2}\approx \allowbreak 0.004$. From formula (\ref{N59}) we
estimate the linear entropy of the steady state as $ L \approx
0.126$ and from (\ref{N58}) we estimate the von Neumann entropy as
$E\approx 0.263$. Surprisingly, these estimations match the exact
values already for relatively low pump intensities. For higher pump
intensities the agreement is even better.

Note that in the strong pump approximation, a very simple relation
holds $|\gamma |  =2|  \delta |$, which gives crude estimations of
$S=2/3$ and $F=2/3$. Since $x=\frac{1
}{3}\sqrt{3}-\frac{1}{2}\approx 0.0774$, the linear entropy is
estimated as $L= 1-\sqrt{3}/2\approx 0.134$ and the entropy as $E
\approx 0.278$. We can also estimate the weight coefficients of
steady state decomposition as $p_{0}\approx 0.928$, $p_{1}\approx
0.067$, $p_{2}\approx 0.005$. So, even the crude estimations well
match the exact values obtained numerically in the previous
section.

In figure 12 we have compared the exact quantum weight factors
$p_{k}$ of the steady state and the approximate Gaussian weight
factors given by (\ref{N56}) with $x\approx 0.072$ calculated from
(\ref{N57}) for the resonator parameters $G=0.2,$ $\gamma _{0}=1$
and $p=5$. This graphical representation shows the validity of our
approximation.
\begin{figure}
\epsfxsize=6cm\centerline{\epsfbox{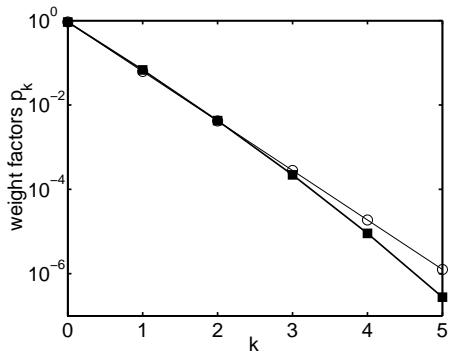}} \caption{Exact
quantum weight factors $p_{k}$ of the steady state (thick curve
with squares) vs the Gaussian state approximation given by
(\ref{N56}) (thin curve with circles)}
\end{figure}

\section{Conclusions}

We studied the evolution of the quantum noise and mixedness of a
dissipative non-linear oscillator, described by a Kerr non-linear
oscillator, pumped by a classical external field. We quantified the
quantum noise by the quadrature squeezing parameter and the Fano
factor, while the mixedness by the quantum von Neumann and linear
entropies. Dissipation was described in the standard
Heisenberg-Langevin approach by coupling the system to
zero-temperature reservoir. We demonstrated that initial pure
states, including the Fock states, coherent states, and their
finite superpositions, exhibited fast decoherence and evolved into
the same steady mixed state being well approximated by a
nonclassical Gaussian state as verified by the Bures metric and the
quantum relative entropy. We presented analytical formulas and
numerical results for the steady state parameters to show that the
state exhibits quadrature noise squeezing and sub-Poissonian
statistics if the thermal reservoir is cold. We observed,
especially for the initial nonclassical states including the Fock
states or the finite superpositions of coherent states, a rapid
increase in the mixedness at the very beginning of the evolution,
and then for longer times a fall in the mixedness to the same low
value of the asymptotic nonclassical Gaussian state in the case of
the cold reservoir.

\section*{Acknowledgements}

The authors thank Profs. Vlasta Pe\v{r}inov\'a, Wies\l{}aw
Leo\'nski, and Ryszard Tana\'{s} for stimulating discussions. This
work was supported by Projects No. J14/98 and No. LN00A015 of the
Czech Ministry of Education and by the EU grant under QIPC,
Project No. IST-1999-13071 (QUICOV).

\appendix

\section{Gaussian states}

The Gaussian state is defined as a state with Gaussian
quasidistribution and Gaussian characteristic functions (see,
e.g., \cite{perina,perinova,paris}), which can be identified
completely by the three parameters
\begin{equation}
\alpha =\langle \hat{a}\rangle ,\;\;\;\;\;B=\langle \hat{a} ^{\dag
}\hat{a}\rangle -\langle \hat{a}^{\dag }\rangle \langle
\hat{a}\rangle,\quad C=\langle \hat{a}^{2}\rangle -\langle
\hat{a}\rangle ^{2}. \label{N54}
\end{equation}
Thus the Gaussian state can be defined by its Cahill-Glauber
$s$-parametrized quasidistribution function
\begin{eqnarray}
{\cal W}^{(s)}(\beta) &=&\frac{1}{\pi \sqrt{K_s}}\exp \Big(
-\frac{1-s+2B}{2K_s} |  \beta -\alpha |  ^{2} \nonumber\\
&&\hspace{1cm} +\left[ \frac{C^{\ast }}{K_s}\left( \beta -\alpha
\right) ^{2}+\mathrm{c.c.}\right] \Big) , \label{N55}
\end{eqnarray}
where $K_s=\left( 1/2-s/2+B\right) ^{2}-|C|^{2}$. The Gaussian
state is a natural generalization of coherent ($B=C=0$) and
chaotic ($C=0$) states. Note that in a special case of $s=1$,  the
Glauber-Sudarshan function $P(\beta) \equiv {\cal W}^{(1)}(\beta)$
exists if $K_{1}>0$, otherwise the state described by (\ref{N55})
is nonclassical, and thus referred to as the NCGS. The Gaussian
states are mixed states with the weight coefficients
\begin{equation}
p_{k}=\frac{x^{k}}{\left( 1+x\right) ^{1+k}},  \label{N56}
\end{equation}
where
\begin{equation}
x=\sqrt{\left( B+\frac{1}{2}\right) ^{2}-|  C| ^{2}}-\frac{ 1}{2}.
\label{N57}
\end{equation}

The von Neumann entropy of the Gaussian state (\ref{N55}) reads as
\cite{perinova}
\begin{equation}
E=-\ln \frac{x^{x}}{\left( 1+x\right) ^{1+x}}  \label{N58}
\end{equation}
Similarly, it is easy to show that the purity parameter is
\begin{equation}
P=(1+2x)^{-1},  \label{N59}
\end{equation}
while the quadrature noise squeezing parameter reads as
\begin{equation}
S=1+2(B-|C|)   \label{N60}
\end{equation}
and its Fano factor is
\begin{equation}
F=1+2B+\frac{C\alpha ^{\ast 2}+C^{\ast }\alpha ^{2}}{| \alpha |
^{2}+B}, \label{N61}
\end{equation}
where the parameters $\alpha$, $B$ and $C$ are defined by
(\ref{N54}). Note that the squeezing occurs if $B<|C|$ or
equivalently if $K_{1}<0$, so any NCGS exhibits quadrature
squeezing.


\end{document}